\documentclass[aps,pra,amssymb,twocolumn,superscriptaddress,10pt,tightenlines,nofootinbib,secnumarabic, graphics,floatfix,nobibnotes]{revtex4-1}

\usepackage{graphicx}
\usepackage{dcolumn}
\usepackage{bm}
\usepackage{color}
\usepackage{subfigure}
\usepackage{url}
\usepackage{hyperref}
\usepackage{epstopdf}
\usepackage{soul}
\usepackage{amsmath} %???

\begin{document}
%NEW COMMANDS
\newcommand{\m}[1] {\left<#1\right>}
\newcommand{\abs}[1] {\lvert#1\rvert}
\renewcommand{\r}{\rho}

\newcommand{\beq}{\begin{equation}}
\newcommand{\eeq}{\end{equation}}
\newcommand{\barr}{\begin{eqnarray}}
\newcommand{\earr}{\end{eqnarray}}

\def\bra#1{\langle{#1}|}
\def\ket#1{|{#1}\rangle}
\def\sinc{\mathop{\text{sinc}}\nolimits}
\def\cV{\mathcal{V}}
\def\cH{\mathcal{H}}
\def\cT{\mathcal{T}}
\def\F{\mathcal{F}}
\def\S{\mathcal{S}}
\def\I{\mathcal{I}}
\def\D{\mathrm{D}}
\def\br{\bm{\rho}}
\def\ee{\mathrm{e}}
\def\ii{\mathrm{i}}
\def\dd{\mathrm{d}}
\def\refo{\mathrm{ref}}
\renewcommand{\Re}{\mathop{\text{Re}}\nolimits}
\newcommand{\tr}{\mathop{\text{Tr}}\nolimits}

\definecolor{dgreen}{rgb}{0,0.5,0}
\newcommand{\green}{\color{dgreen}}
\newcommand{\BLUE}[1]{\text{\color{blue} #1}}
\newcommand{\GREEN}[1]{\textbf{\color{green}#1}}
\newcommand{\REV}[1]{\textbf{\color{red}[[#1]]}}
\newcommand{\KY}[1]{\textbf{\color{dgreen}[[#1]]}}
\newcommand{\rev}[1]{{\color{red}[[#1]]}}

\def\HN#1{{\color{magenta}#1}}
\def\DEL#1{{\color{yellow}#1}}

\title{Signal-to-noise properties of correlation plenoptic imaging with chaotic light}

\author{Giovanni Scala}
\affiliation{Dipartimento Interateneo di Fisica, Universit\`a degli Studi di Bari, I-70126 Bari, Italy}
\affiliation{INFN, Sezione di Bari, I-70125 Bari, Italy}

\author{Milena D'Angelo}
\affiliation{Dipartimento Interateneo di Fisica, Universit\`a degli Studi di Bari, I-70126 Bari, Italy}
\affiliation{Istituto Nazionale di Ottica (INO-CNR), I-50125 Firenze, Italy}
\affiliation{INFN, Sezione di Bari, I-70125 Bari, Italy}

\author{Augusto Garuccio}
\affiliation{Dipartimento Interateneo di Fisica, Universit\`a degli Studi di Bari, I-70126 Bari, Italy}
\affiliation{INFN, Sezione di Bari, I-70125 Bari, Italy}

\author{Saverio Pascazio}
\affiliation{Dipartimento Interateneo di Fisica, Universit\`a degli Studi di Bari, I-70126 Bari, Italy}
\affiliation{INFN, Sezione di Bari, I-70125 Bari, Italy}
\affiliation{Istituto Nazionale di Ottica (INO-CNR), I-50125 Firenze, Italy}

\author{Francesco V. Pepe}\email{francesco.pepe@ba.infn.it}
\affiliation{INFN, Sezione di Bari, I-70125 Bari, Italy}

\begin{abstract}
\noindent 
Correlation Plenoptic Imaging (CPI) is a novel imaging technique, that exploits the correlations between the intensity fluctuations of light to perform the typical tasks of plenoptic imaging (namely, refocusing out-of-focus parts of the scene, extending the depth of field, and performing 3D reconstruction), without entailing a loss of spatial resolution. Here, we consider two different CPI schemes based on chaotic light, both employing ghost imaging: the first one to image the object, the second one to image the focusing element. We characterize their noise properties in terms of the signal-to-noise ratio (SNR) and compare their performances. We find that the SNR can be significantly higher and easier to control in the second CPI scheme, involving standard imaging of the object; under adequate conditions, this scheme enables reducing by one order of magnitude the number of frames for achieving the same SNR.
\end{abstract}

\maketitle

\section{Introduction}

Plenoptic imaging is a recently established optical imaging technique, based on the idea of recording both the spatial distribution and propagation direction of light in a single exposure \cite{adelson}. Although the first feasible proposal to apply plenoptic imaging to digital cameras dates back to the mid-2000s \cite{ng}, the seminal intuition can be attributed to Lippmann \cite{lippmann} one century earlier. Plenoptic imaging is currently employed in a very wide range of applications, that include stereoscopy \cite{adelson,muenzel,levoy}, microscopy \cite{microscopy1,microscopy2,microscopy3,microscopy4}, particle image velocimetry \cite{piv}, particle tracking and sizing \cite{tracking}, and wavefront sensing \cite{thesis_wu,eye,atmosphere1,atmosphere2}. Since plenoptic devices are able to simultaneously acquire 2D images from multiple perspectives, they are considered among the fastest and most promising methods for 3D imaging \cite{3dimaging}, as shown by the very recent use in imaging of animal neuronal activity \cite{microscopy4}, surgical robotics \cite{surgery}, endoscopy \cite{endoscopy} and blood-flow visualization \cite{piv2}. 

Currently available plenoptic imaging devices are based on the intensity measurement on a single detector \cite{ng,georgiev1,georgiev2}. Their key component is a microlens array, that produces multiple images of some reference plane, not coinciding with the object plane defined by the main lens. In this way, the direction of light from the object plane to such reference plane can be traced, enabling to reconstruct (\textit{refocus}) out-of-focus parts of the scene, extend the depth of field, and perform 3D imaging in post-processing. However, capturing directional information entails a fundamental tradeoff with the image resolution. In particular, spatial resolution in plenoptic devices cannot reach the diffraction limit, as determined by the light wavelength and the numerical aperture of the imaging system. 

Several technologies have been developed in the field of quantum imaging, which go beyond the capabilities of standard imaging and interferometry systems \cite{pittman,qu_superres,qu_superres2,sofi,undetected,dangelo_kim,scarcelli_er,tamma,genovese_review}. Recently, a technique named Correlation Plenoptic Imaging (CPI) \cite{cpi_review} has been shown to overcome the typical tradeoff between spatial and directional resolution of plenoptic imaging, by exploiting intensity correlations of either chaotic light \cite{cpi_prl,cpi_qmqm,cpi_jopt,cpi_exp} or entangled photon pairs \cite{cpi_technologies}. The key idea of CPI is to encode information of the image and the direction of light in two distinct sensors: the desired information emerges by evaluating intensity correlations. Since two separate sensors are used, the image resolution can reach the diffraction limit. CPI is inspired by ghost imaging with chaotic and entangled light \cite{pittman,gatti,valencia,scarcelliPRL,bennink,devaux,laserphys,shapiro_review}, with a crucial modification: the ``bucket'' detector, collecting all light that propagates in one optical path in ghost imaging, is replaced by a spatially resolving detector in CPI. The resolution of such detector enables to track the direction of light. 

Though the tradeoff between spatial and directional resolution can be overcome by using CPI instead of traditional plenoptic imaging, the former has the disadvantage of requiring the reconstruction of the source statistics, thus losing the single-shot advantage of standard plenoptic imaging. The signal-to-noise ratio (SNR) improves with the number of frames; however, to aim at performing real-time imaging, the number of acquired frames should be as small as possible. The choice of the optimal frame number is particularly delicate in the case of ghost images with chaotic light, characterized by a well-known tradeoff between resolution and SNR \cite{gatti_coh,erkmen,osullivan,brida_pra}. Ways to mitigate such tradeoff involve image analysis techniques \cite{katz,welsh} and alternative measurement schemes \cite{ferri_dgi}. The objective of this paper is to characterize and compare the SNR in two different CPI schemes based on the properties of chaotic light and designed according to complementary concepts (see Fig.~\ref{fig:setups}): the first one ({\sc setup1}) exploits ghost imaging to obtain the image of the object, and standard imaging to get directional information, while in the second one ({\sc setup2}) the object is imaged by a lens, and ghost imaging is used to obtain directional information. 

In Section II, we will outline the problem and define its general aspects. In Section III, we will derive the results that enable one to determine the optimal number of frames to be acquired to achieve the chosen SNR, given the light properties, the optical distances and the object features. The results obtained in the two setups will be compared and interpreted. In Section IV, we will further discuss the perspectives of this research.

\section{Correlation plenoptic imaging schemes}

\begin{figure}
\centering
\includegraphics[width=\linewidth]{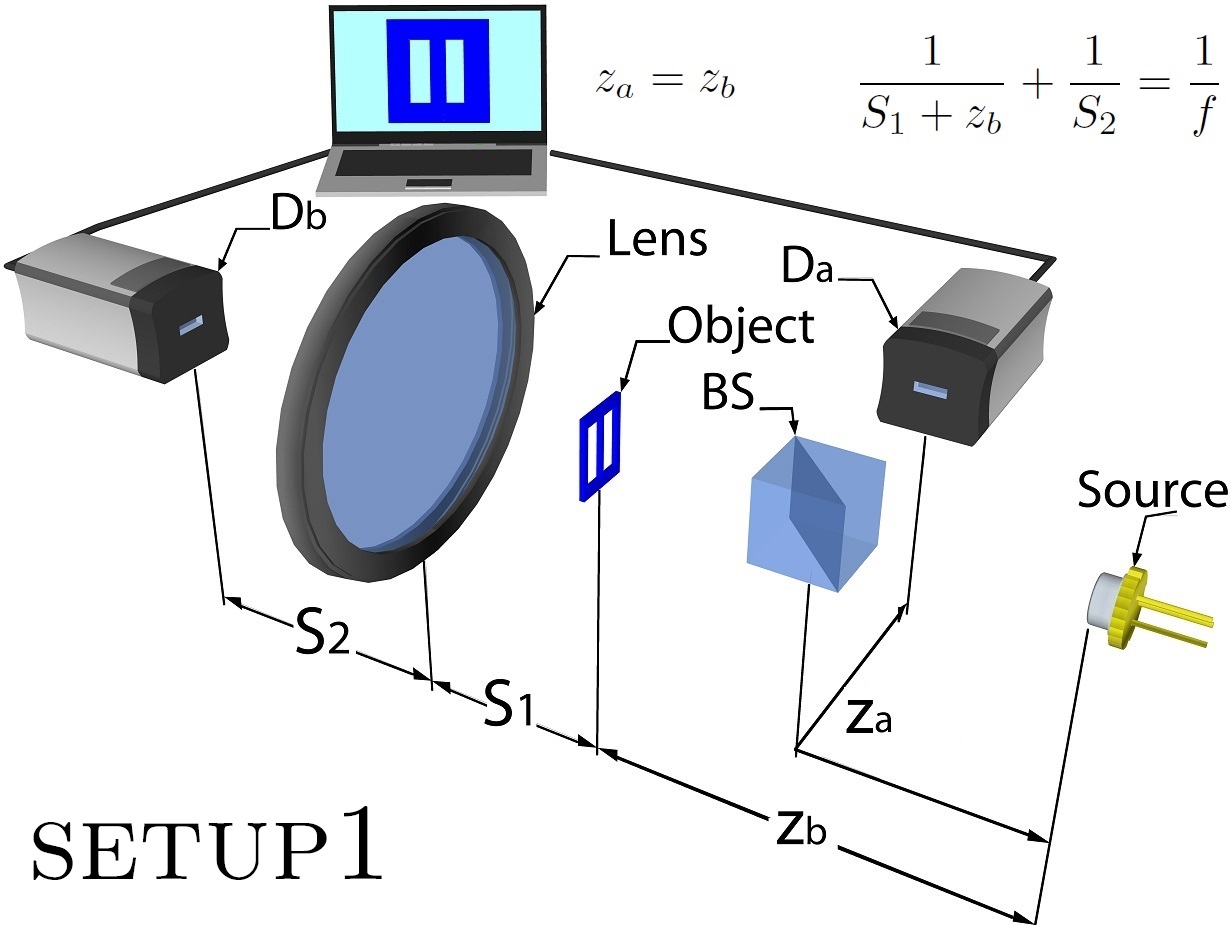} 
\includegraphics[width=\linewidth]{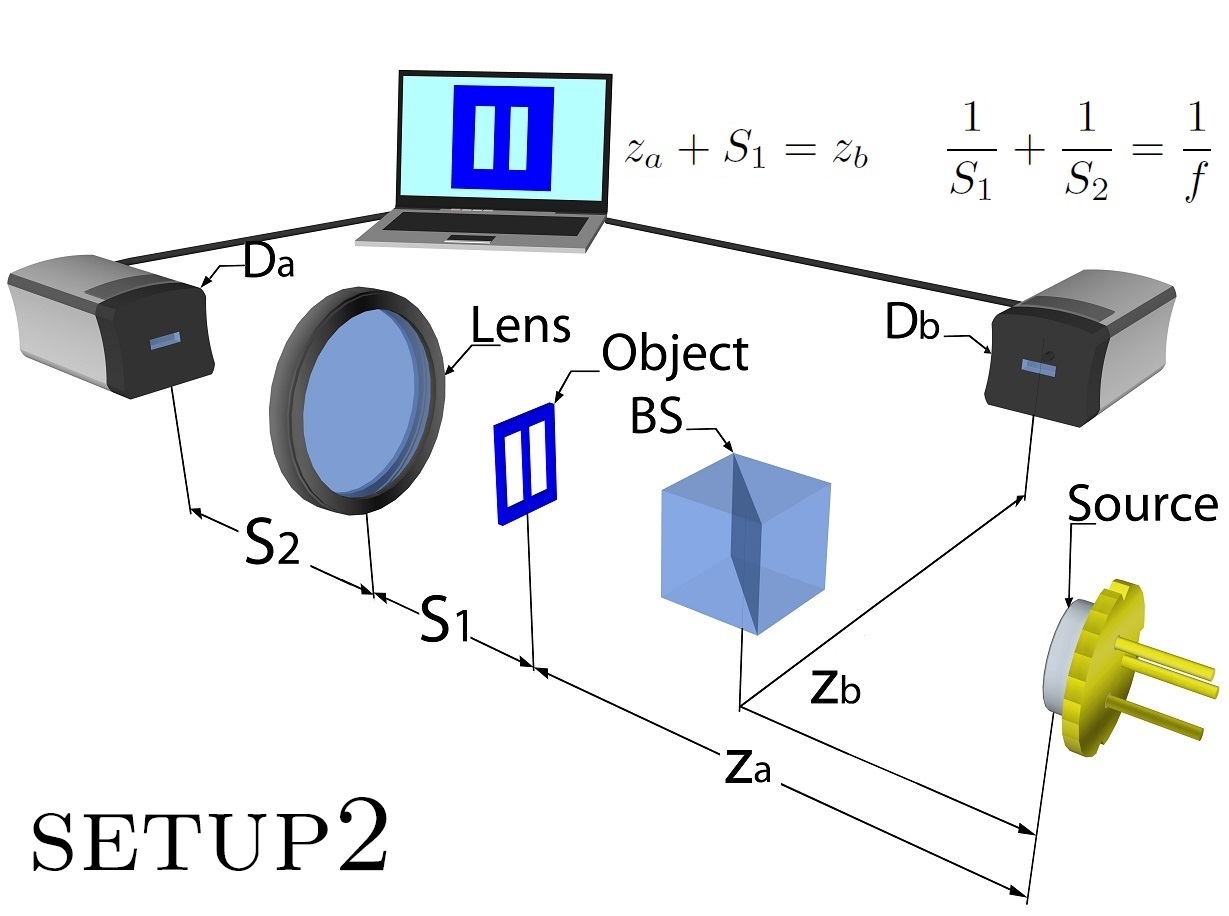}
\caption{Schematic representation of two setups that enable to perform plenoptic imaging by measuring the correlation of intensity fluctuations between points on two spatially resolving detectors $\D_a$ and $\D_b$. Both setups are illuminated by chaotic light, that is split in two paths by a beam splitter, and feature a transmissive object and a lens of focal length $f$. In {\sc setup1} (upper panel), the chaotic source is focused by the lens on detector $\D_b$, while the ``ghost'' image of the object emerges in correspondence of $\D_a$ from the average correlation $\Gamma(\br_a,\br_b)=\m{\Delta i(\br_a) \Delta i(\br_b)}$. {\sc setup2} (lower panel) is based on a different working principle: the image of the object is formed by the lens on $\D_a$, while the ghost image of the lens is retrieved in correspondence of $\D_b$ by computing correlations between $\D_a$ and $\D_b$. In both cases, encoding these two images in the correlation function of Eq.~\eqref{Gamma} provides information on the direction of light in the setup, giving the possibility to recover the image of the object even if the focusing conditions (namely, $z_b=z_a$ for {\sc setup1}, and $1/S_1+1/S_2=1/f$ for {\sc setup2}) are not satisfied.}\label{fig:setups}
\end{figure}

We will consider the two setups ({\sc setup1} and {\sc setup2}) represented in Fig.\ \ref{fig:setups}, for performing correlation plenoptic imaging. These configurations have been proposed in \cite{cpi_prl} and \cite{cpi_jopt}, respectively, and an experimental proof of principle of plenoptic imaging and refocusing in {\sc setup1} has been performed \cite{cpi_exp}. The two schemes essentially differ by the way ghost imaging is employed to obtain an image of either the object plane ({\sc setup1}) or the focusing element ({\sc setup2}). The common feature of the two setups is the fact that light emitted by a chaotic source is split in two paths $a$ and $b$ by a beam splitter (BS), and is recorded at the end of each path by the high-resolution detectors $\D_a$ and $\D_b$. An object is always placed in one of the two paths. More specifically, intensity patterns $I_A(\br_a)$ and $I_B(\br_b)$, with $\br_{a,b}$ the coordinate on each detector plane, are recorded in time to reconstruct the correlation function
\begin{equation}\label{Gamma}
\Gamma_{AB}(\br_a,\br_b)=\m{\Delta I_A(\br_a)\Delta I_{B}(\br_b)},
\end{equation}
with $\Delta I_{A,B}(\br_{a,b}) = I_{A,B}(\br_{a,b})- \m{I_{A,B}(\br_{a,b})}$. The expectation value in \eqref{Gamma} must be evaluated over the source statistics, but it can be approximated by the time average of the product of the intensity fluctuations, provided the source is stationary and ergodic \cite{mandel}. In the discussed setups, the images of the object plane and of the focusing element aperture will be simultaneously encoded in $\Gamma_{AB}(\br_a,\br_b)$.

In {\sc setup1}, an image of the object can be obtained only by measuring intensity correlations between $\D_a$ and $\D_b$. Along path $a$ (the reflected path in figure), light directly impinges on detector $\D_a$, placed at an optical distance $z_a$ from the source. In path $b$ (the transmitted path in figure), a transmissive object lies at a distance $z_b$ from the source. A thin lens of focal length $f$ is placed between the object and the detector $\D_b$, at a distance $S_1$ from the former and $S_2$ from the latter. Such distances are chosen in order to focus the source on $\D_b$ with magnification $M=S_2/(S_1+z_b)$, hence, they satisfy the thin-lens equation $1/S_2+1/(S_1+z_b)=1/f$. In the case $z_b=z_a$, measurement of the correlation function $\Gamma_{AB}(\br_a,\br_b)$ and direct integration over $\br_b$ provides the \textit{focused} ghost image of the object \cite{scarcelliPRL}.

In {\sc setup2}, the image of the lens is recovered from intensity correlations between $\D_a$ and $\D_b$. Along path $b$ (the reflected path in figure), light directly impinges on the detector $\D_b$, placed at an optical distance $z_b$ from the source. In path $a$ (the transmitted path in figure), the transmissive object is placed at a distance $z_a$ from the source. The thin lens of focal length $f$ lies between the object and the detector $\D_a$, at a distance $S_1$ from the former and $S_2$ from the latter. In this case, the setup is designed to obtain a focused ghost image of the lens on the detector $\D_b$: therefore, distances are fixed in order to satisfy $z_b=z_a+S_1$. The object-to-lens and lens-to-$\D_a$ distances are arbitrary. However, it is intuitive that, if $S_2=S_2^f$, such that $1/S_1+1/S_2^f=1/f$, the image of the object will be sharply focused on $\D_a$. 

The refocusing capability of both setups is determined by the fact that the correlation function \eqref{Gamma} encodes multiple coherent images of the object, one for each point $\br_b$ on $\D_b$. The images corresponding to different pixels on $\D_b$ are generally displaced with respect to each other, unless a focusing condition is satisfied. In the focused case, integration over detector $\D_b$ yields an incoherent image. In the out-of-focus cases, the collected coherent images need to be realigned \textit{before} integrating over $\D_b$, following
\begin{equation}\label{Sigmaref}
\Sigma_{\refo} (\br_a) = \m{ \S_{(\alpha,\beta)} (\br_a) } ,
\end{equation}
with
\begin{equation}\label{sigmaref}
\S_{(\alpha,\beta)} (\br_a) = \int \dd^2\br_b \Delta I_A(\alpha\br_a+\beta\br_b) \Delta I_B(\br_b) .
\end{equation}
The parameters $(\alpha,\beta)$, that approach $(1,0)$ at focus, are properly chosen to realign the coherent images depending on the setup, and read 
\begin{equation}\label{refocus}
(\alpha,\beta)= \left\{ \begin{matrix} \displaystyle \left(\frac{z_a}{z_b},-\frac{1}{M}\left(1-\frac{z_a}{z_b}\right)\right) & \text{for {\sc setup1},} \\ & \\  \displaystyle \left(\frac{S_2}{S_2^f},1-\frac{S_2}{S_2^f}\right) & \text{for {\sc setup2}.} \end{matrix}  \right.
\end{equation}
It is evident that, when the focusing conditions are fulfilled, there is no need to shift and rescale the first argument of $\Gamma$, and the high resolution of detector $\D_b$ plays no role. In all other cases, the spatial resolution of $\D_b$ is essential to reconstruct the image of an out-of-focus object, which, by direct integration over $\D_b$, would appear blurred and degraded.

\section{Fluctuations and SNR}

\subsection{General aspects and statistical model}

The objective of this paper is to estimate the signal-to-noise ratio characterizing the refocused images retrieved in {\sc setup1} and {\sc setup2}. To this end, we shall analyze the fluctuations of the refocused observable $\S_{(\alpha,\beta)} (\br_a)$, defined in Eq.~\eqref{sigmaref}, around its average $\Sigma_{\refo}(\br_a)$, namely
\begin{align}\label{F}
\F(\br_a) = & \m{ \S_{(\alpha,\beta)} (\br_a)^2 } - \m{ \S_{(\alpha,\beta)} (\br_a) }^2 \nonumber \\ = & \int \dd^2\br_{b1} \dd^2\br_{b2} \Phi(\br_a,\br_{b1},\br_{b2}),
\end{align}
with $\Phi$ determined by the local fluctuations of the intensity correlations [see Eq.~\eqref{sigmaref}]. Let us assume that $N_f$ frames are collected in time to evaluate the expectation value \eqref{Sigmaref}. Supposing their statistical independence, the root-mean-square error affecting the evaluation of $\Sigma_{\refo} (\br_a)$ can be estimated by $\sqrt{\F(\br_a)/N_f}$. We therefore define the quantity
\begin{equation}\label{SNRgeneral}
R(\br_a) = \sqrt{N_f} \frac{\Sigma_{\refo} (\br_a)}{\sqrt{\F(\br_a)}}.
\end{equation}
as the signal-to-noise ratio. 

A scalar model of the electromagnetic field, in which the effects of polarization are neglected, will be adopted, and we will assume that the radiation emission by the source is an approximately Gaussian random process, stationary and ergodic. In particular, the field $V_S(\br_s)$ at a point $\br_s$ on the source will be characterized by a Gaussian-Schell equal-time correlator \cite{mandel}
\begin{equation}\label{WS}
W_S(\br_s,\br_s') = \m{ V_S(\br_s)V_S^*(\br_s') } = I_s \ee^{- \frac{\br_s^2}{4\sigma_i^2} - \frac{\br_s'^2}{4\sigma_i^2} - \frac{(\br_s-\br_s')^2}{2\sigma_g^2}} ,
\end{equation}
with $I_s$ the peak intensity, $\sigma_i$ the width of the intensity profile $\m{I_S(\br_s)}=W_S(\br_s,\br_s)=I_s\ee^{-\br_s^2/2\sigma_i^2}$, and $\sigma_g$ the transverse coherence length on the source plane. Since we are interested in chaotic sources, characterized by negligible transverse coherence, we will also approximate the mutual coherence function with a delta function
\begin{equation}\label{mutu}
\exp(-\br^2/2\sigma_g^2) \simeq2\pi\sigma_g^2 \delta^{(2)}(\br)
\end{equation}
under the integrals. 

To compute \eqref{Sigmaref} and \eqref{F}, it is necessary to determine up to eight-point field correlators. Using the Gaussian approximation, we will assume that Isserlis-Wick's theorem \cite{isserlis} is valid for the correlators that involve an equal number of $V$'s and $V^*$'s, namely
\begin{equation}\label{Wick}
\m{ \prod_{j=1}^n V_S (\br_j) V_S^*(\br_j')} = \sum_{\mathrm{P}} \prod_{j=1}^n \m{V_S (\br_j) V_S^*(\mathrm{P}\br_j')} ,
\end{equation}
with $\mathrm{P}$ a permutation of the primed indexes, while all other expectation values, including $\m{V}$ and $\m{V^*}$, vanish. Propagation from the source to the detectors along the two paths $a$ and $b$ is deterministic, and depends on the transmission functions of the object and the lens. Concerning propagation in free space, a monochromatic field with frequency $\omega$ and wavenumber $k=\omega/c$, evaluated on a plane at a general longitudinal position $z$, is related to the field at $z_0<z$ by the paraxial transfer function \cite{goodman}:
\begin{equation}\label{free}
V(\br;z) = \frac{-\ii k}{2\pi (z-z_0)} \int \dd^2\br' V(\br';z_0) \ee^{\ii k \left[ \frac{(\br-\br')^2}{2(z-z_0)} + (z-z_0) \right]} .
\end{equation}
The correlators between fields $V_A(\br_a)$ and $V_B(\br_b)$ at the detectors $\D_a$ and $\D_b$, that determine the refocused image $\Sigma_{\refo}$ and the fluctuation $\F$, thus inherit the factorization property \eqref{Wick} from the fields on the source. In particular, since $I_A=V_A^*V_A$ and $I_B=V_B^*V_B$, the correlation of intensity fluctuations between the two detectors, defined in Eq.~\eqref{Gamma}, reads
\begin{equation}
\Gamma_{AB} (\br_a,\br_b) = \left| \m{ V_A(\br_a) V_B^*(\br_b) } \right|^2 .
\end{equation} 
Computation of the fluctuation \eqref{F}, based on the definition \eqref{sigmaref}, also involves the autocorrelations of intensity fluctuations at the same detector, 
\begin{align}
\Gamma_{DD}(\br_{1},\br_{2}) & =\m{ \Delta I_D(\br_{1})\Delta I_D(\br_{2}) } \nonumber \\ & = \left| \m{ V_D(\br_1) V_D^*(\br_2) } \right|^2 ,
\end{align}
with $D=A,B$.

In both setups, $\mathcal{F}(\br_a)$ is determined with good approximation by the contribution that features only the autocorrelations:
\begin{align}\label{F0}
\F_0(\br_a) := \int \dd^2\br_{b1} \dd^2\br_{b2} & \Gamma_{AA}(\alpha\br_a+\beta\br_{b1},\alpha\br_a+\beta\br_{b2}) \nonumber \\ & \times \Gamma_{BB}(\br_{b1},\br_{b2}) .
\end{align}
Other contributions are typically suppressed as 
\begin{equation}
\frac{|\F-\F_0|}{\F_0} \sim \frac{1}{\mathcal{N}_b} ,
\end{equation}
with $\mathcal{N}_b$ the number of transverse modes that propagate towards the detector $\D_b$. Therefore, in the following, we shall approximate $\F\simeq\F_0$ when computing the SNR. However, the full computation of all contributions to $\F$ is presented in the Appendix.

\subsection{Analysis of {\sc setup1}}

Let us first consider {\sc setup1} (Fig.~\ref{fig:setups}, upper panel). Let us call $A(\br)$ the aperture function of the transmissive object, and neglect the finite pupil size of the lens, by assuming that it does not affect propagation along path $b$. Combining free propagation \eqref{free} with transmission through the object and the lens \cite{goodman}, and applying the statistical assumptions \eqref{WS}-\eqref{mutu}-\eqref{Wick} on the field correlations at the source, we obtain the correlation between the fluctuations of the intensities $I_A(\br_a)=|V_A(\br_a)|^2$ and $I_B(\br_b)=|V_B(\br_b)|^2$, which reads
\begin{align}\label{GammaAB1}
\Gamma_{AB}(\br_a,\br_b) & = |S_{AB}|^4 K_{AB} \nonumber \\
& \times  \left| \int \dd^2\br_o A(\br_o) \ee^{-\gamma_a \left( \frac{\br_a}{\alpha} - \br_o \right)^2 -\ii \gamma_b\br_b\cdot\br_o} \right|^2 ,
\end{align} 
with $\alpha= z_a/z_b$ as in \eqref{refocus}, and the coefficients 
\begin{equation}\label{const11}
\gamma_a= \frac{k^2 S_{AB}^2}{2 z_b^2}, \quad \frac{1}{S_{AB}^2} = \frac{1}{\sigma_i^2} + \ii k \left( \frac{1}{z_a}-\frac{1}{z_b} \right), \quad \gamma_b = \frac{k}{Mz_b},
\end{equation}
while $K_{AB}=K_A K_B$, with
\begin{equation}\label{const21}
K_A= I_s \left( \frac{k\sigma_g}{z_a} \right)^2 , \quad K_B = I_s \left( \frac{k^2 \sigma_g}{2\pi M z_b^2} \right)^2 . 
\end{equation}
Since $\gamma_a$ is a generally complex quantity, it will be useful in the following to split it into its real and imaginary parts as $\gamma_a=\gamma_r+\ii \gamma_i$. The result (\ref{GammaAB1}) shows that, by varying $\br_b$, a collection of coherent images of the object is obtained on $\D_a$.

Combining Eq.~\eqref{GammaAB1} with the definitions \eqref{Sigmaref}--\eqref{refocus}, we determine the refocused image 
\begin{equation}\label{Sigmaref1}
\Sigma_{\refo} (\br_a) = \frac{\pi}{2\delta^2 \gamma_r} |S_{AB}|^4 K_{AB} I_{\Sigma}(\br_a),
\end{equation}
with
\begin{align}\label{ISigma1}
I_{\Sigma}(\br_a) & = \int \dd^2\br_1 \dd^2\br_2 A^*(\br_a-\br_1) A(\br_a-\br_2) \ee^{\ii \frac{\gamma_b(\br_1^2-\br_2^2)}{2\delta}} \nonumber \\
& \times \exp\left[ -\left(\frac{\gamma _r}{2}+\frac{(\gamma _b-2 \delta \gamma_i)^2}{8 \delta^2 \gamma_r}\right)(\br_1-\br_2)^2 \right]
\end{align}
and $\delta=\beta/\alpha=(1-z_b/z_a)/M$. This quantity is regular in the focused limit $\delta\to 0$, where the $\br_a$-dependent part of the integral takes the form
\begin{equation}\label{Sigmafocused1}
\left. I_{\Sigma}(\br_a) \right|_{z_b=z_a} \sim \int \dd^2\br |A(\br_a-\br)|^2 \exp\left(-\frac{\br^2}{\sigma_A^2}\right) ,
\end{equation} 
which is exactly the unit-magnification incoherent image obtained in the case of lensless ghost imaging \cite{scarcelliPRL,osullivan}, whose point-spread function is determined by the squared Fourier transform of the source intensity profile. In the geometrical optics limit ($k\to\infty$), the dominant contribution to the integral \eqref{ISigma1} comes from the stationary point of the real and imaginary parts of the exponent, yielding
\begin{equation}\label{Sigmaref1g}
\Sigma_{\refo}^{(g)} (\br_a) = I_s^2 \frac{\pi \sigma_g^4}{\sigma_A^2} |A(\br_a)|^2, 
\end{equation}
which also shows that $\Sigma_{\refo}$ actually provides a refocused image of the object, characterized by unit magnification. 

Let us now evaluate the autocorrelations of the intensity fluctuations
\begin{align}
\Gamma_{AA}(\br_{a1},\br_{a2}) & = \sigma_i^4 K_{A}^2 \exp \left( - \frac{(\br_{a1}-\br_{a2})^2}{\sigma_A^2} \right) , \label{GammaAA1} \\
\Gamma_{BB}(\br_{b1},\br_{b2}) & = \sigma_i^4 K_{B}^2 \Biggl| \int \dd^2\br_1 \dd^2 \br_2 A^*(\br_1) A(\br_2)  \nonumber \\
&  \times \ee^{ - \frac{(\br_{1}-\br_{2})^2}{2\sigma_B^2} - \ii \gamma_b (\br_{b2}\cdot\br_2 - \br_{b1}\cdot\br_1) } \Biggr|^2 , \label{GammaBB1}
\end{align}
where
\begin{equation}
\sigma_D = \frac{z_d}{k\sigma_i},
\end{equation}
with $D=A,B$, is the transverse coherence length on the planes at a distance $z_{d}=z_a,z_b$ from the source. The correlation functions \eqref{GammaAA1}-\eqref{GammaBB1} enable to evaluate the dominant contribution to the variance of the correlation of the intensity fluctuations in Eq.~\eqref{F0}, that reads
\begin{equation}\label{F01}
\F_0(\br_a) = (2\pi)^3 \left( K_{AB} \frac{\sigma_A^2 \sigma_i^4}{\gamma_b \beta} \right)^2 I_{\F_0} ,
\end{equation}
with
\begin{align}
I_{\F_0} = \int & \Bigl(\prod_{j=1}^3 \dd^2\br_j\Bigr) A(\br_1)A^*(\br_2)A(\br_3)A^*(\br_1+\br_3-\br_2) \nonumber \\
& \times \exp\left( - \frac{(\br_2-\br_3)^2}{\sigma_B^2} - \frac{(\br_1-\br_2)^2}{2\sigma_i^2(1-z_a/z_b)} \right) .
\end{align}
The most relevant (and interesting) feature of such quantity is its independence on the coordinate $\br_a$ on $\D_a$. Therefore, the signal $\Sigma_{\refo}(\br_a)$ is noisy and superposed to a further background noise. Such constant background noise stems from the fact that the intensity profile of the light impinging on $\D_a$ is, in the case of {\sc setup1}, not related to the spatial profile of the signal $\Sigma_{\refo}(\br_a)$: actually, as one can easily check, the intensity profile $\m{I_A}$ on $\D_a$ is approximately uniform, and carries no information on the object transmission function $|A(\br_a)|^2$.

At this point, the SNR can be exactly evaluated, as a function of the number of collected frames, by using the results \eqref{Sigmaref1}-\eqref{F01} in the expression \eqref{SNRgeneral}. A useful and intuitive estimate is given by the geometrical-optics approximation of $R(\br_a)$, which reads
\begin{equation}\label{R1g}
R^{(g)}(\br_a) = \sqrt{2} \pi \sigma_B \sigma_i \left|1-\frac{z_b}{z_a}\right| |A(\br_a)|^2 \sqrt{ \frac{N_f}{J^{(g)}} } ,
\end{equation}
with
\begin{equation}\label{J1g}
J^{(g)} = \int \dd^2\br_1 \dd^2\br_2 |A(\br_1)A(\br_2)|^2 \ee^{ - \frac{(\br_1-\br_2)^2}{2\sigma_i^2(1-z_a/z_b)^2} } .
\end{equation}
Let us first discuss this result in the focused case, in which $z_a=z_b$ and $\sigma_A=\sigma_B$. The integrand of \eqref{J1g} becomes localized around $\br_1=\br_2$, and the value of the SNR reduces to
\begin{equation}\label{R1gfocused}
\left. R^{(g)}(\br_a) \right|_{z_b=z_a} = \sqrt{ N_f \frac{ \pi \sigma_B^2 }{ \int\dd^2\br |A(\br)|^4 } } |A(\br_a)|^2 .
\end{equation}
The above expression highlights the dependence of the SNR on the ratio between the coherence area $\sim\sigma_B^2$ on the object and an ``effective area'' of the object itself, given by the integral of the $|A|^4$ factor, which is equal to the actual area in the case of binary transmission function. Since the same coherence area determines the resolution through \eqref{Sigmafocused1}, this result entails the well-known tradeoff between resolution and SNR typical of ghost imaging \cite{gatti_coh,erkmen,brida_pra,osullivan}. 

\begin{figure}
\centering
\includegraphics[width=\linewidth]{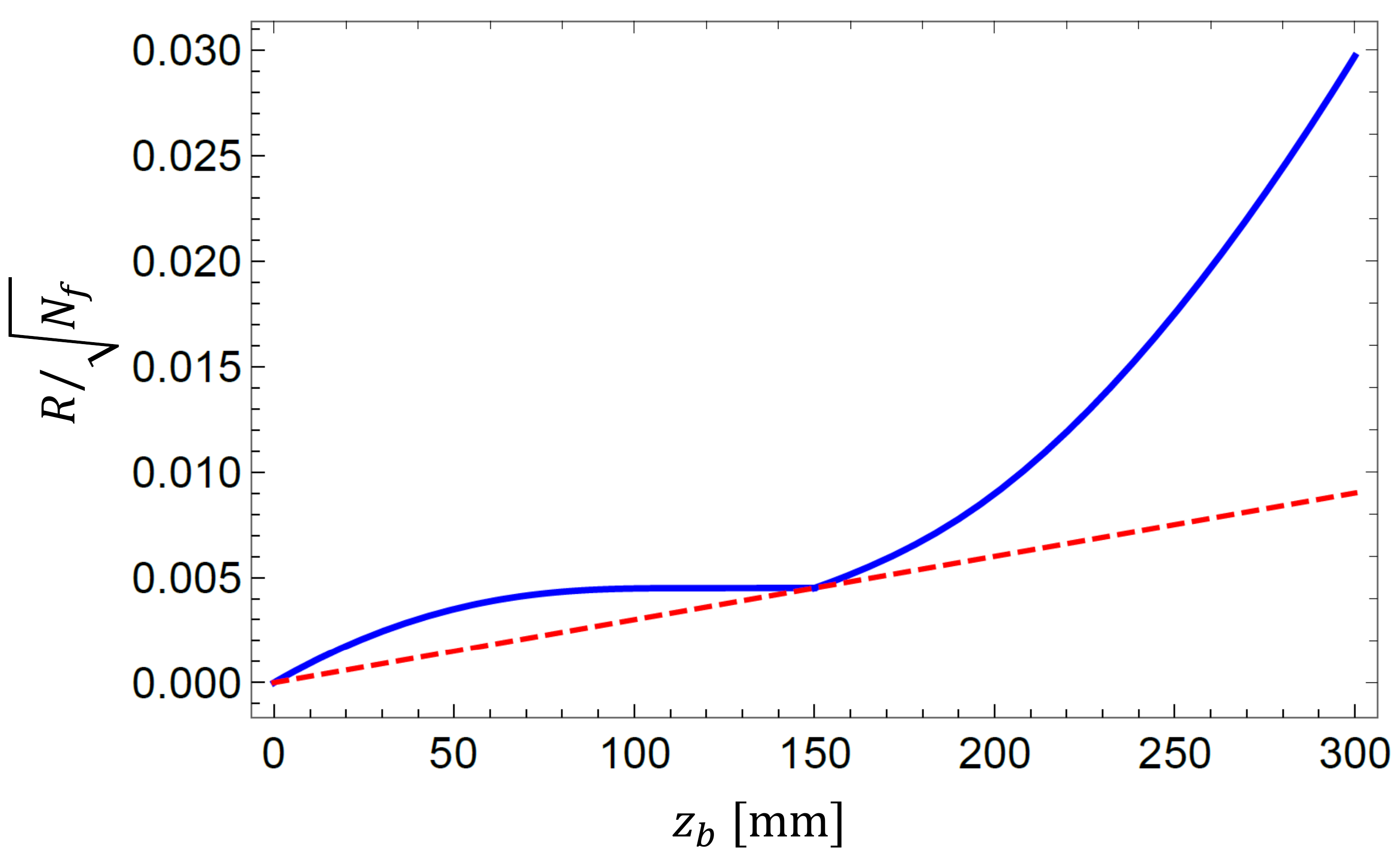}
\caption{Signal-to-noise ratio, normalized to the square root of the number of frames, for the refocused image \eqref{Sigmaref1} obtained in {\sc setup1} (solid blue line). The source is characterized by wavelength $\lambda=532\,\mathrm{nm}$ and a Gaussian intensity profile of width $\sigma_i=2.5\,\mathrm{mm}$, and is placed at a fixed distance $z_a=150\,\mathrm{mm}$ from detector $\D_a$. The focused image, obtained at $z_b=z_a$, is characterized by resolution $\Delta x=10\,\mu\mathrm{m}$. The values are computed in correspondence of a totally transmissive point ($A=1$) of a binary object with transmissive area $A_{\mathrm{obj}}=4\,\mathrm{mm}^2$. The SNR for a ghost image taken at $z_a=z_b$, as a function of $z_b$ (red dashed line), is shown for comparison.}\label{fig:snrs1}
\end{figure}

Deep in the out-of-focus regime, when $\sigma_i|1-z_b/z_a|$ becomes larger than the typical size of the object, the exponential modulation under the integral \eqref{J1g} can be neglected, yielding
\begin{equation}\label{R1gnf}
R^{(g)}(\br_a) \simeq \sqrt{\frac{N_f}{2}} \lambda z_b \left|1-\frac{z_b}{z_a}\right| \frac{|A(\br_a)|^2}{\int\dd^2\br |A(\br)|^2 }  ,
\end{equation}
with $\lambda=2\pi/k$ the light wavelength. This expression shows a less trivial dependence on the longitudinal position $z_b$ of the refocused plane, but can still be interpreted in terms of the resolution-SNR tradeoff. Actually, as discussed in \cite{cpi_prl,cpi_exp}, a good estimate of the resolution of the refocused image is given by $\Delta x = (\lambda z_b/a) |1-z_b/z_a|$, where $a$ is the typical linear size of the smallest transmissive parts of the object. Notice, however, that the inverse dependence on the effective area of the object has changed with respect to the focused case\eqref{R1gfocused}. As a rule of thumb, we can estimate the SNR of refocused images as
\begin{equation}
\frac{R^{(g)}(\br_a)}{\sqrt{N_f}} \sim \sqrt{ \frac{a^2}{A_{\mathrm{obj}}} } \sqrt{ \frac{(\Delta x)^2}{A_{\mathrm{obj}}} } |A(\br_a)|^2 ,
\end{equation}
a result that depends on the product of the ratios (resolution cell)/(total area) and (smallest detail area)/(total area). In Fig.\ \ref{fig:snrs1}, we show the behavior of the SNR in {\sc setup1} as a function of the source-to-object distance $z_b$, comparing the result with the case of a focused ghost image taken with $z_a=z_b$ [Eq.\ \eqref{R1gfocused}]. The higher SNR of correlation plenoptic imaging is related to the lower resolution of the refocused image with respect to the focused ghost image.

\subsection{Analysis of {\sc setup2}}

In the analysis of {\sc setup2}, we shall consider a finite-size pupil $P(\br)$ of the lens, that determines the spatial resolution, and assume that the transverse size of the source is asymptotically large, namely, that the finite size of the source does not affect in a relevant way the correlation of the intensity fluctuations. In this way, the most relevant part of the mutual correlation function reads
\begin{align}\label{GammaAB2}
\Gamma_{AB}(\br_a,\br_b) = & \left( \frac{z_a (z_a+S_1)}{k S_1} \right)^2 K_{AB}  \nonumber \\
& \times \Bigl| \int \dd^2\br_o A(\br_o) \tilde{P}_{\beta}\Bigl( \frac{\br_o}{S_1} + \frac{\br_a}{S_2} \Bigr)  \ee^{\frac{\ii k}{S_1} \br_o\cdot\br_b} \Bigr|^2 ,
\end{align}
where
\begin{equation}
\tilde{P}_{\beta}(\bm{q}) = \int \dd^2\br_{\ell} P(\br_{\ell}) \exp \left( \frac{\ii k \beta}{2 S_2} \br_{\ell}^2 - \ii k \bm{q}\cdot\br_{\ell} \right),
\end{equation}
which coincides with the Fourier transform of the pupil function, represents the coherent PSF of the focused image (obtained when $S_2=S_2^f=(1/f-1/S_1)^{-1}$ and $\beta=0$), and $K_{AB}=K_A K_B$, with 
\begin{equation}\label{KAB2}
K_A = I_s \left( \frac{k^3 \sigma_g}{(2\pi)^2 S_1 S_2 z_a} \right)^2, \quad K_B = I_s \left( \frac{k \sigma_g}{z_a+S_1} \right)^2 .
\end{equation}
The refocused image, defined by $(\alpha,\beta)$ in Eq.\ \eqref{refocus} (second formula), reads
\begin{equation}\label{Sigmaref2}
\Sigma_{\refo} (\br_a) = \left( \frac{2\pi z_a (z_a+S_1)}{k^2} \right)^2 K_{AB} I_{\Sigma}(\br_a),
\end{equation}
with
\begin{align}
I_{\Sigma}& (\br_a) = \int \dd^2\br_1 \dd^2\br_2 \dd^2 \br_o P^*(\br_1) P(\br_2) \ee^{ - \frac{\ii k \beta}{2 S_2} (\br_2-\br_1)^2 } \nonumber \\
& \times A^*(\br_o) A\Bigl( \br_o + \frac{\beta S_1}{S_2} (\br_2-\br_1) \Bigr) \ee^{ \ii k \Bigl( \frac{\br_b}{S_2^f} + \frac{\br_o}{S_1} \Bigr)\cdot (\br_2-\br_1) } .
\end{align}
One can easily check that, in the focused case $\beta=0$, the above expression reduces to the incoherent image of the transmission function of the object, whose point-spread function is related to the usual square modulus of the Fourier transform of the lens pupil function $P$. In the general case, the plenoptic imaging property emerges when considering the geometrical optics limit, in which the complicated expression \eqref{Sigmaref2} simplifies to
\begin{equation}
\Sigma_{\refo}^{(g)} (\br_a) = I_s^2 \frac{k^2 \sigma_g^4}{S_2^2} A_{\mathrm{lens}} \left| A\left(- \frac{\br_a}{\mu} \right) \right|^2  ,
\end{equation}
where $\mu=S_2^f/S_1$ is the absolute magnification provided by the lens, and $A_{\mathrm{lens}}=\int\dd^2\br |P(\br)|^2$ is the (effective) area of the lens.

The autocorrelations, computed in the same regime of large source width $\sigma_i$, read
\begin{align}
\Gamma_{AA}(\br_{a1},\br_{a2}) & = \Bigl( 2\pi K_A \frac{z_a^2}{k^2} \Bigr)^2  \Biggl| \int \dd^2\br_o |A(\br_o)|^2   \nonumber \\ & \times\tilde{P}_{\beta}\Bigl( \frac{\br_o}{S_1} + \frac{\br_{a2}}{S_2} \Bigr) \tilde{P}_{\beta}^* \Bigl( \frac{\br_o}{S_1} + \frac{\br_{a1}}{S_2} \Bigr) \Biggr|^2 , \label{GammaAA2} \\
\Gamma_{BB}(\br_{b1},\br_{b2}) & = \pi \sigma_i^2 K_B^2 \left( \frac{z_a+S_1}{k} \right)^2 \delta^{(2)} (\br_{b1}-\br_{b2}) . \label{GammaBB2}
\end{align}
Notice that the autocorrelation on the detector $\D_b$ diverges with increasing $\sigma_i$: therefore, even if the finite size of the source can become irrelevant for the average correlation of intensity fluctuations, the variance of this quantity crucially depends on it.

Also in this case, the dominant contribution to the variance of $\S_{(\alpha,\beta)}$ can be evaluated by considering the autocorrelations. However, the computation of \eqref{F0} must take into account the finite size of the detector $\D_b$, since, integrating without bounds on $\br_b$ would yield a divergent result. Since the role of $\D_b$ is to detect the ghost image of the lens, which is characterized by unit magnification, the optimal size of this detector is given by the size of the lens. Following these considerations and the result \eqref{GammaAA2}, valid in the limit of large source width $\sigma_i$, one obtains
\begin{align}\label{F02}
\F_0(\br_a) = & \pi \sigma_i^2 K_B^2 \left( \frac{z_a+S_1}{k} \right)^2 \nonumber \\ & \times \int_{\D_b} \dd^2\br_b \Gamma_{AA}(\alpha\br_a+\beta\br_b,\alpha\br_a+\beta\br_b) .
\end{align}
In the focused case, the integral in \eqref{F02} is trivially proportional to $\Gamma_{AA}(\br_a,\br_a)$, which is $\br_a$-dependent, as opposed to the case of {\sc setup1}. In the general case, the analytic evaluation of \eqref{F02} can become impossible when considering the actual detector area as the integration domain. However, one can perform the computation by regularizing the integral with a Gaussian envelope function $\exp(-\pi\br_b^2/A_{\D_b})$, where $A_{\D_b}$ is the area of detector $\D_b$ (or, better, the area of the part of the detector that accommodates the image of the lens). 

The geometrical-optics approximation of \eqref{F02} reads
\begin{equation}\label{F02g}
\F_0^{(g)}(\br_a) = \frac{A_{\D_b}}{4\pi} \left( (2\pi)^4 \frac{(z_a+S_1)\sigma_i z_a^2 S_2^2}{k^5 (1-S_2/S_2^f)^2} \right)^2 J^{(g)}(\br_a) ,
\end{equation}
with 
\begin{align}\label{J2g}
J^{(g)}(\br_a) & = \Biggl( \int  \dd^2\br_o |A(\br_o)|^2 \nonumber \\ & \times \left| P \Biggl( \frac{S_2/S_1}{1-S_2/S_2^f} \Bigl(\br_o+\frac{\br_a}{\mu}\Bigr)\Biggr) \right|^2 \Biggr)^2.
\end{align}
The spatial behavior of the variance $\F_0$ is now much less trivial than the constant behavior found in {\sc setup1}. Actually, in the focused limit $S_2\to S_2^f$, the integrand of \eqref{J2g} becomes infinitely localized around $\br_o=-\br_a/\mu$, leading to $J^{(g)}(\br_a)\propto |A(-\br_a/\mu)|^4$. This means that, at least in the focused case, \textit{noise is proportional to the signal}. Such feature is related to the fact that, in this case, the field transmitted by the object is focused on $\D_a$, and this is reflected in all the correlation functions involving $\D_a$. This feature is absent in the focused case of {\sc setup1}, in which the field impinging on $\D_a$ extends well beyond the shape of the object, and the image emerges only from intensity correlation measurements. In the opposite limit of large defocusing, instead, the spatial dependence of the lens pupil function $P$ under the integral \eqref{J2g} becomes irrelevant, and the result $J^{(g)}(\br_a)\propto (\int\dd^2\br|A(\br)|^2)^2$ implies that the measurement of $\Sigma_{\refo}(\br_a)$ comes with a uniform background noise. However, as we shall presently find, such background is more easily controllable than the one surrounding the ghost image in {\sc setup1}. 

Based on the above considerations, the estimate of the SNR based on Eq.~\eqref{SNRgeneral} is less trivial than in {\sc setup1}, since the denominator depends on $\br_a$ and shows different spatial behaviors with varying defocusing. The geometrical-optics expression of the SNR reads
\begin{align}\label{R2g}
R^{(g)}(\br_a) = & 2\sigma_B \sqrt{ N_f \frac{\pi}{A_{\D_b} J^{(g)}(\br_a)} } \left( \frac{1-S_2/S_2^f}{S_2/S_1} \right)^2 \nonumber \\ & \times \left| A\left(- \frac{\br_a}{\mu} \right) \right|^2 \int \dd^2\br |P(\br)|^2 .
\end{align}
In the focused case, the above quantity reduces to the simple expression
\begin{equation}\label{R2gfocused}
\left. R^{(g)}(\br_a) \right|_{S_2=S_2^f} = 2\sigma_B \sqrt{N_f \frac{\pi}{A_{\D_b}}} ,
\end{equation}
a result that does not depend on $\br_a$, since noise is proportional to the signal. The constant SNR in \eqref{R2gfocused} is essentially the square root of the ratio of the coherence area $\sim\sigma_B^2$ on $\D_b$ and the area $A_{\D_b}$ of the same detector, which can also be interpreted as (coherence area on the lens)/(area of the lens), in perfect analogy with Eq.~\eqref{R1gfocused}, after replacing the object with the lens.  The SNR thus coincides with the one expected for the ghost image of the lens. 

\begin{figure}
\centering
\includegraphics[width=\linewidth]{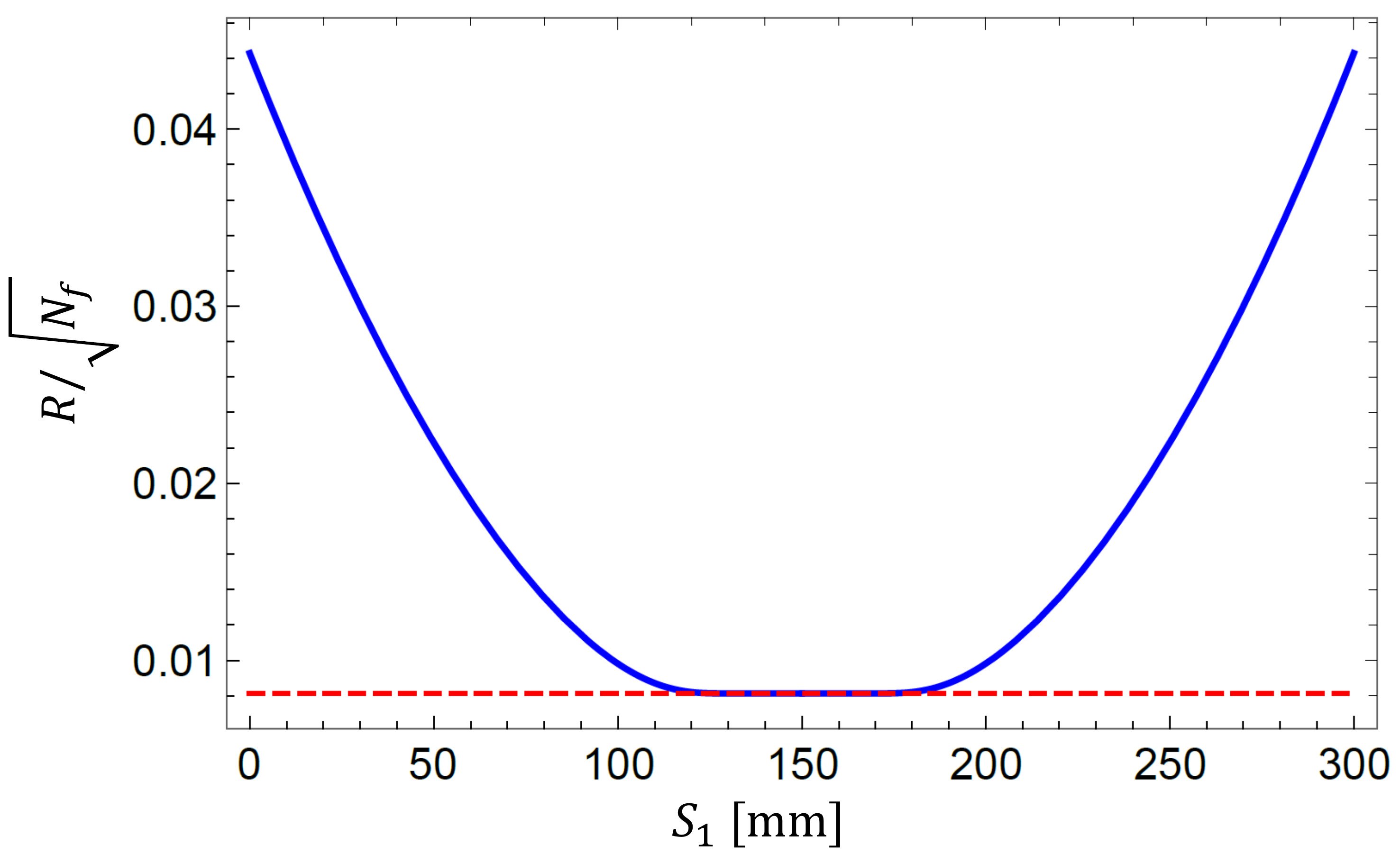}
\caption{Signal-to-noise ratio, normalized to the square root of the number of frames, for the refocused image \eqref{Sigmaref2} obtained in {\sc setup2} (solid blue line). The source is characterized by a wavelength $\lambda=532\,\mathrm{nm}$ and a Gaussian intensity profile of width $\sigma_i=2.5\,\mathrm{mm}$, and is placed at a fixed distance $z_b=z_a+S_1=300\,\mathrm{mm}$ from the detector $\D_b$. The lens has a focal length $f=75\,\mathrm{mm}$ and a Gaussian pupil function of width $\sigma_p=2.5\,\mathrm{mm}$. Fixing the value $S_2=2f=150\,\mathrm{mm}$, the focused image, obtained at $S_1=S_2$, is characterized by the same resolution, depth of field and magnification as in the case shown in Fig.~\ref{fig:snrs1}. The values are computed in correspondence of a totally transmissive point ($A=1$) of a binary object with transmissive area $A_{\mathrm{obj}}=4\,\mathrm{mm}^2$. The SNR for a ghost image taken at $S_2=S_2^f=(1/f-1/S_1)^{-1}$ as a function of $S_1$ (red dashed line) is shown for comparison.}\label{fig:snrs2}
\end{figure}

In the out-of-focus case, a background noise emerges, and the SNR becomes similar in form to \eqref{R1gnf}, yielding
\begin{align}\label{R2gnf}
R(\br_a)^{(g)} \simeq & 2\sigma_B \sqrt{N_f \frac{\pi}{A_{\D_b}}} \left( \frac{1-S_2/S_2^f}{S_2/S_1} \right)^2 \nonumber \\ & \times \frac{\int\dd^2\br |P(\br)|^2}{\int\dd^2\br |A(\br)|^2} |P(0)|^2 \left| A\left(- \frac{\br_a}{\mu} \right) \right|^2 .
\end{align}
The ratio between the area of the lens and the area of the object is generally large for image magnification $\mu\gtrsim 1$, and the SNR also increases quadratically with defocusing, providing a generally more favorable picture compared to {\sc setup1}. A good rule to estimate the order of magnitude of the refocused image SNR thus reads
\begin{equation}
\frac{R^{(g)}(\br_a)}{\sqrt{N_f}} \sim \Bigl( \frac{S_2/S_1}{1-S_2/S_2^f} \Bigr)^2 \sqrt{\frac{\sigma_B^2}{A_{\mathrm{lens}}}} \frac{A_{\mathrm{lens}}}{A_{\mathrm{obj}}} \left| A\left(- \frac{\br_a}{\mu} \right) \right|^2 ,
\end{equation} 
where we have assumed that the area of the detector is matched to the area of the lens. In Fig.\ \ref{fig:snrs2}, we represent the behavior of the SNR in {\sc setup2} as a function of the object-to-lens distance $S_1$, and compare the result with the case of a focused image taken with $S_2^f=S_2$ (notice that $S_2^f$ is a function of $S_1$).

\subsection{Summary of the results}

We have discussed the properties of the signal-to-noise ratio for {\sc setup1} and {\sc setup2}, finding that the results obtained for the latter are generally more advantageous than for the former. In the focused case, {\sc setup2} is characterized by the suppression of background noise, that, on the other hand, is a typical feature affecting the ghost image obtained in {\sc setup1}. Moreover, noise in {\sc setup1} increases with improving resolution on the object, thus entailing a trade-off between resolution and SNR trade-off. In the out-of-focus case, background noise is present in both configurations. However, in {\sc setup1} it depends on small quantities, namely the ratios $(\Delta x)^2/A_{\mathrm{obj}}$ between the area of an effective resolution cell and the total area of the object, and $a^2/A_{\mathrm{obj}}$, where the numerator is the area corresponding to the size $a$ of the finest details of the object. In {\sc setup2}, instead, we find that the SNR depends also on the ratio $A_{\mathrm{lens}}/A_{\mathrm{obj}}$, a quantity that is not necessarily small. Therefore, we expect that a smaller number of frames is needed to achieve the same resolution in {\sc setup2} compared to {\sc setup1}. 

To get a quantitative hint of the SNR improvement in {\sc setup2}, we compare the results shown in Figs.~\ref{fig:snrs1}-\ref{fig:snrs2}, which are referred to two cases that are as homogeneous as possible in terms of resolution, depth of field and magnification of the focused image. We find that he ratio between the SNR in {\sc setup2} and {\sc setup1} at fixed $N_f$ is consistently larger than one: when such a ratio reaches values around $3.2$, for an object placed at $z_b=S_1=80\,\mathrm{mm}$, one tenth of the frames is needed in {\sc setup2} to reach the same SNR as in {\sc setup1}. Notice that, by considering the expressions \eqref{R1gnf}-\eqref{R2gnf}, the ratio of the SNRs for out-of-focus images is very weakly dependent of the light wavelength and the area of the object, provided the conditions for the validity of geometrical optics approximation are satisfied.

\section{Conclusions and outlook}

Performing plenoptic imaging by correlation measurements has the potential to improve 3D imaging and microscopy, since it combines high resolution with the possibility to gain directional information. The results obtained in this Article provide the experimenter with rules to determine the scaling of the SNR with the number of frames, and consequently to fix the number of frames needed for a fast and accurate imaging of the scene. The problem of optimizing the acquisition time is particularly relevant both in view of real-time imaging and in all those cases in which additional difficulties in retrieving intensity correlations are present, as it happens when considering unconventional sources like X rays \cite{pelliccia,schneider} to perform CPI. In our future research, we plan to extend our analysis to the case in which CPI is performed with entangled photons \cite{cpi_technologies}, investigating whether the remarkable results observed in other kinds of imaging schemes \cite{brida_nat,meda,samantaray}, in which the shot-noise limit can be overcome, would yield analogous improvements in a setup oriented to plenoptic imaging.

\section*{Acknowledgments}
MD, AG and FVP are supported by Istituto Nazionale di Fisica Nucleare (INFN) through the project ``PICS''. MD, AG, SP and GS are supported by Istituto Nazionale di Fisica Nucleare (INFN) through the project ``QUANTUM''. AG and MD are supported by the Italian Ministry of Education, University and Research (MIUR) through project PON Ricerca e Innovazione ARS01\_00141.

\appendix
\section{}

We have identified and discussed the term $\F_0$, defined in Eq.~\eqref{F0}, as the most relevant term to determine the SNR in both setups. Here, we provide the computation of the remaining terms characterizing the local fluctuations $\F(\br_a)$: 
\begin{equation}\label{DeltaF}
\Delta\F(\br_a)=\F(\br_a)-\F_0(\br_a)= \sum_{j=1}^{7} \F_j(\br_a)
\end{equation}
that characterize the variance at a point $\br_a$, in the geometrical-optics approximation. In \eqref{DeltaF}, all but one term are conjugate to each other, namely $\F_j(\br_a)=\F_{3+j}^*(\br_a)$ with $2\leq j\leq 4$. The independent contributions read
\begin{align}
\F_1 (\bm{\r} _a)=\int & \dd^2\bm{\r} _{\text{b1}}\dd^2\bm{\r} _{\text{b2}}
\Gamma _{AB}\left(\alpha \bm{\r}_a+\beta \bm{\r} _{\text{b1}},\bm{\r} _{\text{b2}}\right) \nonumber \\
& \times
\Gamma _{AB}\left(\alpha \bm{\r}_a+\beta \bm{\r} _{\text{b2}},\bm{\r} _{\text{b1}}\right), \label{F1} \\
\F_2(\bm{\r} _a) =
\int & \dd^2\bm{\r}_{b_1}\dd^2\bm{\r}_{b_2} W_{AB}(\alpha\bm{\r}_a+\beta\bm{\r}_{b_1},\bm{\r}_{b_1}) \nonumber \\ 
& \times W_{AB}(\alpha\bm{\r}_a+\beta\bm{\r}_{b_2},\bm{\r}_{b_2}) \nonumber \\ 
& \times W^*_{AB}(\alpha\bm{\r}_a+\beta\bm{\r}_{b_2},\bm{\r}_{b_1}) \nonumber \\ 
& \times W^*_{AB}(\alpha\bm{\r}_a+\beta\bm{\r}_{b_1},\bm{\r}_{b_2}) \label{F2}
\end{align}
and
\begin{align}\label{F34}
\F_{3(4)} (\bm{\r}_a)=
\int &
\dd^2\bm{\r} _{\text{b1}} \dd^2\bm{\r} _{\text{b2}}
W_{BB}\left(\bm{\r} _{\text{b1}},\bm{\r} _{\text{b2}}\right) \nonumber \\
& \times W_{AA}\left(\alpha \bm{\r} _a+\beta \bm{\r} _{\text{b2}},\alpha \bm{\r} _a+\beta \bm{\r} _{\text{b1}}\right) \nonumber \\
& \times W_{AB}\left(\alpha \bm{\r} _a+\beta \bm{\r} _{\text{b1(b2)}},\bm{\r} _{\text{b1}}\right) \nonumber \\
& \times W_{AB}^*\left(\alpha \bm{\r} _a+\beta \bm{\r} _{\text{b2(b1)}},\bm{\r} _{\text{b2}}\right) .
\end{align}
Notice that, in the focused case, both $\F_1$ and $\F_2$ exactly reduce to the squared refocus image $\Sigma_{\refo}^2$, defined in Eq.~\eqref{Sigmaref}.

\subsection{Results for {\sc setup1}}

The term $\F_1$ reads
\begin{align}\label{F11}
\F_1 (\bm{\r} _a)=&
\Big(\frac{\pi}{2\delta^2 \gamma_r }\Big)^2
\lvert S_{AB} \rvert^8
K_{AB}^{2}
\I_{1}(\bm{\r}_a),
\end{align}
with
\begin{align}
\I_{1}(\bm{\r}_a)=&
\int
\mathcal{A}_1(\bm{\r}_a;\{\bm{\r}_i\})
\ee^{s_1(\{\bm{\r}_i\})}
\prod_{i=1}^{4}\dd^2\bm{\r}_i,
\end{align}
where $\delta=(1-z_b/z_a)/M$, $S_{AB}$, $K_{AB}$ and $\gamma_r=\mathrm{Re} \gamma_a$ are defined in Eqs.~\eqref{const11}-\eqref{const21}, and	
\begin{align}\label{A1}
\mathcal{A}_1(\bm{\r}_a;\{\bm{\r}_i\})=&
A^*(\bm{\r}_a-\bm{\r}_1)
A(\bm{\r}_a-\bm{\r}_2) \nonumber \\ 
& \times
A^*(\bm{\r}_a-\bm{\r}_3)
A(\bm{\r}_a-\bm{\r}_4), 
\end{align}
\begin{align}
s_1(\{\bm{\r}_i\})=&
-\gamma_a(\bm{\r}_2^2+\bm{\r}_4^2)
-\gamma_a^*(\bm{\r}_1^2+\bm{\r}_3^2) \nonumber
\\&+
\frac{R_1^2(\{\bm{\r}_i\})+R_2^2(\{\bm{\r}_i\})}{8 \delta^2 \gamma_r}. \label{apps11}
\end{align}
The functions $R_1$ and $R_2$ in \eqref{apps11} are defined as follows:
\begin{align}
R_1(\{\bm{\r}_i\}) = &R(\bm{\r}_1,\bm{\r}_2,\bm{\r}_3,\bm{\r}_4) = \ii \gamma_b(\bm{\r}_4-\bm{\r}_3)  \nonumber\\
 &-2\gamma_a\delta\bm{\r}_2-2\gamma_a^*\delta\bm{\r}_1,\label{eqr1} \\
\label{eqr2}
R_2(\{\bm{\r}_i\}) = &R(\bm{\r}_3,\bm{\r}_4,\bm{\r}_2,\bm{\r}_1).
\end{align}
In the geometrical optics limit, the function $\F_1$ in \eqref{F11} reduces to
\begin{equation}\label{F11g}
\F_1^{(g)} (\bm{\r} _a)=
\frac{4 \pi ^6}{\gamma _b^4 \gamma _r^2}
\lvert S_{AB}\rvert^8 K_{AB}^2
\lvert A(\bm{\r}_a)\rvert^4.
\end{equation}
	
The term $ \F_2 $ reads
\begin{align}\label{F21}
\F_2 (\bm{\r} _b)=&
\Big(\frac{\pi}{2\delta^2 \gamma_r }\Big)^2
\lvert S_{AB} \rvert^8
K_{AB}^2
\I_{2}(\bm{\r}_a),
\end{align}
with
\begin{align}
\I_{2}(\bm{\r}_a)=&
\int
\mathcal{A}_1(\bm{\r}_a;\{\bm{\r}_i\})
\ee^{s_2(\{\bm{\r}_i\})}
\prod_{i=1}^{4}\dd^2\bm{\r}_i.
\end{align}
and $ \mathcal{A}_1 $ defined in Eq.~\eqref{A1}, while the argument of the exponential reads
\begin{align}
s_2(\{\bm{\r}_i\})=&
-\gamma_a(\bm{\r}_2^2+\bm{\r}_1^2)
-\gamma_a^*(\bm{\r}_4^2+\bm{\r}_3^2) \nonumber
\\&+
\frac{R_3^2(\{\bm{\r}_i\})+R_4^2(\{\bm{\r}_i\})}{8 \delta^2 \gamma_r}, 
\end{align}
with [see Eq.~\eqref{eqr1}]
\begin{align}\label{eqr3}
R_3(\{\bm{\r}_i\})=R(\bm{\r}_4,\bm{\r}_1,\bm{\r}_3,\bm{\r}_1),\\
\label{eqr4}
R_4(\{\bm{\r}_i\})=R(\bm{\r}_3,\bm{\r}_2,\bm{\r}_4,\bm{\r}_2).
\end{align}
The geometrical optics result
\begin{equation}\label{F21g}
\F_2^{(g)} (\bm{\r} _a)=
\frac{4 \pi ^6}{\gamma _b^3\gamma_r\left(\gamma _b \gamma _r+4 \ii \delta|\gamma _a|^2\right)}
\lvert S_{AB}\rvert^8 K_{AB} ^2
\lvert A(\bm{\r}_a)\rvert^4 .
	\end{equation}

The terms $\F_3$ and $\F_4$ in Eq.~\eqref{F34} are both complex and very similar to each other. In this case,
\begin{align}\label{F341}
\F_{3(4)} (\bm{\r} _b)=&
\frac{\pi^2}{q_1 q_2}
\lvert S_{AB} \rvert^4
\sigma_i^4
K_{AB}^2
\I_{3(4)}(\bm{\r}_a),
\end{align}
with
\begin{align}
\I_{3(4)}(\bm{\r}_a)=&
\int
\mathcal{A}_3(\bm{\r}_a;\{\bm{\r}_i\})
\ee^{s_{3(4)}(\{\bm{\r}_i\})}
\prod_{i=1}^{4}\dd^2\bm{\r}_i,
\end{align}
where
\begin{align}\label{A3}
\mathcal{A}_{3}(\br_a;\{\bm{\r}_i\})=&
A^*(\bm{\r} _a- \bm{\r} _{\text{1}})
A(\bm{\r} _a- \bm{\r} _{\text{2}}) \nonumber \\
& \times
A(\bm{\r} _a- \bm{\r} _{\text{3}})	
A^*(\bm{\r} _a- \bm{\r} _{\text{4}}) ,
\end{align}	
coincides, for an object characterized by a real transmission function, with $\mathcal{A}_{1}$ defined in Eq.~\eqref{A1}, and 
\begin{align}
s_{3(4)} (\{ \bm{\r} _i \}) = &
-\gamma _a \bm{\r} _3^2
-\gamma _a^* \bm{\r} _4^2
-
\frac{\left( \bm{\r} _1- \bm{\r} _2\right){}^2}{2 \sigma _B^2} + \frac{T_{3(4)}(\{\bm{\r}_{i}\})^2}{4 q_1} \nonumber
\\&+
\frac{1}{4 q_2}
\left[
U_{3(4)}(\{\bm{\r}_{i}\}) + \frac{\beta^2}{2q_1 \sigma_A^2} T_{3(4)}(\{\bm{\r}_{i}\})
\right]^2, \label{apps341}
\end{align}
with
\begin{align}
T_{3(4)}(\{\bm{\r}_{i}\}) 
	=&-2\delta \gamma_a \bm{\r}_{3}+\ii \gamma_b(\bm{\r}_{3(2)}-\bm{\r}_{1(4)}),\\
U_{3(4)}(\{\bm{\r}_{i}\})
=&-2\delta \gamma_a^* \bm{\r}_{4}-i \gamma_b(\bm{\r}_{4(1)}-\Delta\bm{\r}_{3}).	
\end{align}
The coefficients $q_1$ and $q_2$ appearing in Eq.~\eqref{apps341} are defined as
\begin{equation}\label{q}
q_1=\frac{\beta^2}{2\sigma_A^2}+\gamma_a\delta^2,
\qquad
q_2=q_1^*-\frac{\beta^4}{4q_1\sigma_A^4}.
\end{equation}
In the geometrical optics limit, $\F_3$ and $\F_4$ approach the same value, namely
\begin{align}\label{F341g}
\F_{3}^{(g)} (\bm{\r} _a) & = \F_{4}^{(g)} (\bm{\r} _a) = \lvert S_{AB}\rvert^8 K_{AB}^2 \lvert A(\bm{\r}_a)\rvert^4 \nonumber \\
&	\times \frac{64 i \pi ^6 \sigma _A^2 \sigma _B^2 \left(2 \delta ^2 \gamma _a \sigma _A^2+\beta ^2\right)}{\gamma _b^2\left(
\ii \beta ^4|\gamma _a|^2-8 \delta ^2 \gamma _a \sigma _A^4 \gamma _v-\left(2 \beta \sigma _A\right){}^2 \gamma _u \right)}
\end{align}
where 
\begin{equation}
\gamma _v=(2\left|\gamma _a|^2\delta -\ii \gamma _b\gamma _r\right)\gamma _b ,
\quad
\gamma _u= \gamma _v+2 \ii \delta ^2 \gamma _a^2 \gamma _a^* .
\end{equation}

\subsection{Results for {\sc setup2}}

As in the main text, we will consider a lens with a finite pupil function $P(\br)$ and an asymptotically large source.
The term $\F_1$, as defined in \eqref{F1}, reads
\begin{align}\label{F12}
\F_1 (\bm{\r}_a)&=
\left(
\frac{2\pi z_a (z_a+S_1)}{k^2}
\right)^4K_{AB}^2 
I_{1}(\bm{\r}_a), 
\end{align}
with
\begin{align}
I_{1}(\bm{\r}_a)&=
\int 
\dd^2\bm{\r}_{o1}\dd^2\bm{\r}_{o2}
\prod _{j=1}^4 \dd^2\bm{\r} _{j}
\mathcal{A}_1(\br_{o1},\br_{o2},\{\bm{\r}_{i}\}) \nonumber \\ & \times
\mathcal{P}_1(\{\bm{\r}_{i}\}) \exp\Biggl\{
\ii k
\Biggl[
\frac{\beta}{2 S_2}(\bm{\r}_{2}^2-\bm{\r}_{1}^2+\bm{\r}_{4}^2-\bm{\r}_{3}^2)
\nonumber 
\\ & +\frac{\bm{\r}_{o1}}{S_1}(\bm{\r}_{1}-\bm{\r}_{2}) +
\frac{\bm{\r}_{o2}}{S_1}(\bm{\r}_{3}-\bm{\r}_{4}) \nonumber \\ &
+\frac{\bm{\r}_a}{S_2^f}(\bm{\r}_{1}-\bm{\r}_{2}+\bm{\r}_{3}-\bm{\r}_{4}) \nonumber 
\\ & -
\frac{\beta}{S_2}\bigl(
\bm{\r}_{2}(\bm{\r}_{4}-\bm{\r}_{3})
+\bm{\r}_{4}(\bm{\r}_{2}-\bm{\r}_{1})
\bigr)\Biggr]\Biggr\} ,
\end{align}
with $K_{AB}=K_A K_B$ as defined in Eq.~\eqref{KAB2}, $\beta=1-S_2/S_2^f$, and
\begin{align}
\mathcal{A}_1(\br_{o1},\br_{o2},\{\bm{\r}_{i}\})& =	
A^*(\bm{\r}_{o1})
A\left(\bm{\r}_{o1}+\frac{S_1\beta}{S_2}(\bm{\r}_{4}-\bm{\r}_{3})\right) \nonumber \\	
\times &
A^*(\bm{\r}_{o2})
A\left(\bm{\r}_{o2}+\frac{S_1\beta}{S_2}(\bm{\r}_{2}-\bm{\r}_{1})\right), \label{A12} \\	
\mathcal{P}_1(\{\bm{\r}_{i}\})&=
P^*(\bm{\r}_{1})
P(\bm{\r}_{2})
P^*(\bm{\r}_{3})
P(\bm{\r}_{4}). \label{P12}
\end{align}
The stationary-phase approximation provides the result
\begin{align}\label{F12g}
\F_1^{(g)} (\bm{\r} _a)=&
\left(\frac{S_1 S_2K_{AB}}{\beta}
\left(\frac{4\pi^2 z_a (z_a+S_1)}{k^3}\right)^2 \right)^2 \nonumber
\\\times & \int \dd^2\br_o \dd^2\br_{\ell}
\left| A\left(-\bm{\r}_o-2\frac{\bm{\r}_a}{\mu}\right)\right|^2
\left| A\left(\bm{\r}_o\right)\right|^2	 \nonumber
\\\times& \left| P\left(\bm{\r}_{\ell}\right)\right|^2	
\left| P\left(\bm{\r}_{\ell}-\frac{S_2}{S_1\beta}\left(\bm{\r}_o+\frac{\bm{\r}_a}{\mu}\right)\right)\right|^2.
\end{align}
with $\mu=S_2^f/S_1$.

The result for $\F_2$ is 
\begin{align}\label{F22}
\F_2 (\bm{\r}_a)&=
\left(
\frac{2\pi z_a (z_a+S_1)}{k^2}
\right)^4K_{AB}^2 
I_{2}(\bm{\r}_a), 
\end{align}
with
\begin{align}
I_{2}(\bm{\r}_a)&=
\int 
\dd^2\bm{\r}_{o1}\dd^2\bm{\r}_{o2}
\prod _{j=1}^4 \dd^2\bm{\r} _{j}
\mathcal{A}_2(\br_{o1},\br_{o2},\{\bm{\r}_{i}\}) \nonumber \\ & \times
\mathcal{P}_2(\{\bm{\r}_{i}\}) \exp\Biggl\{
\ii k
\Biggl[
\frac{\beta}{2 S_2}(\bm{\r}_{1}^2+\bm{\r}_{2}^2-\bm{\r}_{3}^2-\bm{\r}_{4}^2)
\nonumber 
\\ & +\frac{\bm{\r}_{o2}}{S_1}(\bm{\r}_{4}-\bm{\r}_{1}) +
\frac{\bm{\r}_{o1}}{S_1}(\bm{\r}_{3}-\bm{\r}_{2}) \nonumber \\ &
+\frac{\bm{\r}_a}{S_2^f}(\bm{\r}_{3}+\bm{\r}_{4}-\bm{\r}_{2}-\bm{\r}_{1}) \nonumber 
\\ & -
\frac{\beta}{S_2}\bigl(
\bm{\r}_{1}(\bm{\r}_{3}-\bm{\r}_{1})
+\bm{\r}_{2}(\bm{\r}_{4}-\bm{\r}_{2})
\bigr)\Biggr]\Biggr\} ,
\end{align}
where
\begin{align}
\mathcal{A}_2(\br_{o1},\br_{o2},\{\bm{\r}_{i}\})& =	
A^*(\bm{\r}_{o1})
A\left(\bm{\r}_{o1}+\frac{S_1\beta}{S_2}(\bm{\r}_{2}-\bm{\r}_{4})\right) \nonumber \\	
\times &
A^*(\bm{\r}_{o2})
A\left(\bm{\r}_{o2}+\frac{S_1\beta}{S_2}(\bm{\r}_{4}-\bm{\r}_{3})\right), \label{A22} \\	
\mathcal{P}_2(\{\bm{\r}_{i}\})&=
P(\bm{\r}_{1})
P(\bm{\r}_{2})
P^*(\bm{\r}_{3})
P^*(\bm{\r}_{4}). \label{P22}
\end{align}
The geometrical optics approximation of \eqref{F22} reads
\begin{align}\label{F22g}
\F_2^{(g)} & (\bm{\r} _a)=
\left(\frac{S_1 S_2 K_{AB}}{\beta}
\left(\frac{4\pi^2 z_a (z_a+S_1)}{k^3}\right)^2 \right)^2 \nonumber
\\ \times & \left| A\left(-\frac{\bm{\r}_a}{\mu}\right)\right|^2 \int \dd^2\br_o \dd^2\br_{\ell}
\ee^{-\ii \frac{k S_2}{S_1^2 \beta} \left(\br_o+\frac{\br_a}{\mu}\right)^2} 
\left| A\left(\bm{\r}_o\right)\right|^2	 \nonumber
\\ \times& \left| P\left(\bm{\r}_{\ell}\right)\right|^2	
\left| P\left(\bm{\r}_{\ell}-\frac{S_2}{S_1\beta}\left(\bm{\r}_o+\frac{\bm{\r}_a}{\mu}\right)\right)\right|^2.
\end{align}

Let us finally compute the remaining independent terms $\F_3$ and $\F_4$. Since, after the assumption of asymptotically large source, $W_{BB}(\br_{b1},\br_{b2})\propto \delta(\br_{b1}-\br_{b2})$, the two terms are incidentally equal to each other, and read
\begin{align}\label{F342}
\F_3 (\bm{\r}_a)&= \F_4 (\bm{\r}_a) =
\left(
\frac{2\pi z_a (z_a+S_1)}{k^2}
\right)^4 K_{AB}^2 
I_{3}(\bm{\r}_a),
\end{align}
with
\begin{align}
I_{3}(\bm{\r}_a)&=
\int 
\dd^2\bm{\r}_{o1}\dd^2\bm{\r}_{o2}
\prod _{j=1}^4 \dd^2\bm{\r} _{j}
\mathcal{A}_3(\br_{o1},\br_{o2},\{\bm{\r}_{i}\}) \nonumber \\ & \times
\mathcal{P}_1(\{\bm{\r}_{i}\}) \exp\Biggl\{
\ii k
\Biggl[
\frac{\beta}{2 S_2}(\bm{\r}_{2}^2-\bm{\r}_{1}^2+\bm{\r}_{4}^2-\bm{\r}_{3}^2)
\nonumber 
\\ & +\frac{\bm{\r}_{o1}}{S_1}(\bm{\r}_{1}-\bm{\r}_{2}) +
\frac{\bm{\r}_{o2}}{S_1}(\bm{\r}_{3}-\bm{\r}_{4}) \nonumber \\ &
+\frac{\bm{\r}_a}{S_2^f}(\bm{\r}_{3}-\bm{\r}_{4}-\bm{\r}_{2}+\bm{\r}_{1}) \nonumber 
\\ & -
\frac{\beta}{S_2}\bigl(
\bm{\r}_{2}(\bm{\r}_{3}-\bm{\r}_{4}-\bm{\r}_{2}-\bm{\r}_{1})
\bigr)\Biggr]\Biggr\} ,
\end{align}
with $\mathcal{P}_1$ defined in Eq.~\eqref{P12} and
\begin{align}\label{A32}
\nonumber
\mathcal{A}_3(\br_{o1},\br_{o2},\{\bm{\r}_{j}\})&=
A^*(\bm{\r}_{o1})|A(\bm{\r}_{o2})|^2
\\ \times&
A\left(\bm{\r}_{o1}-\frac{\beta S_1}{S_2}(\bm{\r}_{3}-\bm{\r}_{4}-\bm{\r}_{2}+\bm{\r}_{1})\right).
\end{align}
The geometrical optics limit yields
\begin{align}\label{F342g}
\F_3^{(g)} & (\bm{\r}_a)=
\left(\frac{S_1 S_2 K_{AB}}{\beta}
\left(\frac{4\pi^2 z_a (z_a+S_1)}{k^3}\right)^2 \right)^2 \nonumber
\\ \times & \left| A\left(-\frac{\bm{\r}_a}{\mu}\right)\right|^2 \int \dd^2\br_o \dd^2\br_{\ell}
\left| A\left(\bm{\r}_o\right)\right|^2	\left| P\left(\bm{\r}_{\ell}\right)\right|^2 \nonumber
\\ \times& 	
\left| P\left(\bm{\r}_{\ell}-\frac{S_2}{S_1\beta}\left(\bm{\r}_o+\frac{\bm{\r}_a}{\mu}\right)\right)\right|^2.
\end{align}


\begin{thebibliography}{99}

%PLENOPTIC CAMERA

\bibitem{adelson}
E.~H. Adelson and J.~Y. Wang, ``Single lens stereo with a plenoptic camera,'' IEEE Trans. Pattern Anal. Mach. Intell. \textbf{14}, 99 (1992).

\bibitem{ng}
R.~Ng, M.~Levoy, M.~Br{\'e}dif, G.~Duval, M.~Horowitz, and P.~Hanrahan, ``Light field photography with a hand-held plenoptic camera,'' Stanford University Computer Science Tech Report CSTR 2005-02, 2005.

\bibitem{lippmann}
G. Lippmann, ``\'Epreuves r\'eversibles donnant la sensation du relief,'' J. Phys. Theor. Appl. \textbf{7}, 821 (1908).

\bibitem{muenzel}
S.~Muenzel and J.~W. Fleischer, ``Enhancing layered 3d displays with a lens,'' Appl. Opt. \textbf{52}, D97 (2013).

\bibitem{levoy}
M.~Levoy and P.~Hanrahan, ``Light field rendering,'' in \textit{Proceedings of the 23rd Annual Conference on Computer Graphics and Interactive Techniques} (Association for Computing Machinery, New York, 1996), pp. 31–42.

\bibitem{microscopy1}
M.~Levoy, R.~Ng, A.~Adams, M.~Footer, and M.~Horowitz, ``Light field microscopy,'' ACM Trans. Graph. \textbf{25}, 924 (2006).

\bibitem{microscopy2}
M.~Broxton, L.~Grosenick, S.~Yang, N.~Cohen, A.~Andalman, K.~Deisseroth, and M.~Levoy, ``Wave optics theory and 3-d deconvolution for the light field microscope,'' Opt. Express \textbf{21}, 25418 (2013).

\bibitem{microscopy3}
W.~Glastre, O.~Hugon, O.~Jacquin, H.~G. de~Chatellus, and E.~Lacot, ``Demonstration of a plenoptic microscope based on laser optical feedback imaging,'' Opt. Express \textbf{21}, 7294 (2013).

\bibitem{microscopy4}
R.~Prevedel, Y.-G. Yoon, M.~Hoffmann, N.~Pak, G.~Wetzstein, S.~Kato, T.~Schr{\"o}del, R.~Raskar, M.~Zimmer, E.~S. Boyden, and A. Vaziri, ``Simultaneous whole-animal 3d imaging of neuronal activity using light-field microscopy,'' Nat. Methods \textbf{11}, 727 (2014).

\bibitem{piv}
T.~W. Fahringer, K.~P. Lynch, and B.~S. Thurow, ``Volumetric particle image velocimetry with a single plenoptic camera,'' Meas. Sci. Technol. \textbf{26}, 115201 (2015).

\bibitem{tracking}
E.~M. Hall, B.~S. Thurow, and D.~R. Guildenbecher, ``Comparison of three-dimensional particle tracking and sizing using plenoptic imaging and digital in-line holography,'' Appl. Opt. \textbf{55}, 6410 (2016).

\bibitem{thesis_wu}
C.~W. Wu, {\em The plenoptic sensor}, Ph.D. thesis, University of Maryland, College Park, 2016.

\bibitem{eye}
Y.~Lv, R.~Wang, H.~Ma, X.~Zhang, Y.~Ning, and X.~Xu, ``SU-G-IeP4-09: Method of human eye aberration measurement using plenoptic camera over large field of view,'' Med. Phys. \textbf{43}, 3679 (2016).

\bibitem{atmosphere1}
C.~Wu, J.~Ko, and C.~C. Davis, ``Using a plenoptic sensor to reconstruct vortex phase structures,'' Opt. Lett. \textbf{41}, 3169 (2016).

\bibitem{atmosphere2}
C.~Wu, J.~Ko, and C.~C. Davis, ``Imaging through strong turbulence with a light field approach,'' Opt. Express \textbf{24}, 11975 (2016).

\bibitem{3dimaging}
X.~Xiao, B.~Javidi, M.~Martinez-Corral, and A.~Stern, ``Advances in three-dimensional integral imaging: sensing, display, and applications,'' Appl. Opt. \textbf{52}, 546 (2013).

\bibitem{surgery}
A.~Shademan, R.~S. Decker, J.~Opfermann, S.~Leonard, P.~C. Kim, and A.~Krieger, ``Plenoptic cameras in surgical robotics: Calibration, registration, and evaluation,'' in \textit{Proceedings of the 2016 IEEE International Conference on Robotics and Automation (ICRA)} (IEEE, New York, 2016), pp. 708–714.

\bibitem{endoscopy}
H.~N. Le, R.~Decker, J.~Opferman, P.~Kim, A.~Krieger, and J.~U. Kang, ``3-d endoscopic imaging using plenoptic camera,'' in \textit{CLEO: Applications and Technology} (Optical Society of America, Washington, DC, 2016), paper AW4O.2.

\bibitem{piv2}
M.~F. Carlsohn, A.~Kemmling, A.~Petersen, and L.~Wietzke, ``3d real-time visualization of blood flow in cerebral aneurysms by light field particle image velocimetry,'' Proc. SPIE, \textbf{9897}, 989703 (2016).

\bibitem{georgiev1}
T. G. Georgiev and A. Lumsdaine, ``Focused plenoptic camera and rendering,'' J. Electron. Imaging \textbf{19}, 021106 (2010).

\bibitem{georgiev2}
T. Georgiev and A. Lumsdaine, ``The multifocus plenoptic camera,'' Proc. SPIE \textbf{8299}, 829908 (2012).

\bibitem{pittman}
T. B. Pittman, Y. H. Shih, D. V. Strekalov, and A. V. Sergienko, ``Optical imaging by means of two-photon quantum entanglement,'' Phys. Rev. A \textbf{52}, R3429 (1995).

\bibitem{genovese_review}
M. Genovese, ``Real applications of quantum imaging,'' J. Opt. \textbf{18}, 073002 (2016).

\bibitem{qu_superres}
O. Schwartz, J. M. Levitt, R. Tenne, S. Itzhakov, Z. Deutsch, and D. Oron, ``Superresolution Microscopy with Quantum Emitters,'' Nano Lett. \textbf{13}, 5832 (2013).

\bibitem{qu_superres2}
Y. Israel, R. Tenne, D. Oron and Y. Silberberg, ``Quantum correlation enhanced super-resolution localization microscopy enabled by a fibre bundle camera,'' Nat. Commun. \textbf{8}, 14786 (2017) 

\bibitem{sofi}
T. Dertinger, R. Colyer, G. Iyer, S. Weiss, J. Enderlein, ``Fast, background-free, 3D super-resolution optical fluctuation imaging (SOFI),'' PNAS \textbf{106}, 22287 (2009).

\bibitem{undetected}
G. Barreto Lemos, V. Borish, G. D. Cole, S. Ramelow, R. Lapkiewicz, and A. Zeilinger, ``Quantum imaging with undetected photons,'' Nature \textbf{512}, 409 (2014).

\bibitem{dangelo_kim}
M. D'Angelo, Y.H. Kim, S. P. Kulik, and Y. Shih, ``Identifying Entanglement Using Quantum Ghost Interference and Imaging", Phys. Rev. Lett. \textbf{92}, 233601 (2004).

\bibitem{scarcelli_er}
G. Scarcelli, Y. Zhou, and Y. Shih, ``Random delayed-choice quantum eraser via two-photon imaging", Eur. Phys. J. D \textbf{44}, 167 (2007).

\bibitem{tamma}
M. D'Angelo, A. Mazzilli, F. V. Pepe, A. Garuccio, and V. Tamma, ``Characterization of two distant double-slits by chaotic light second-order interference,'' Sci. Rep. \textbf{7}, 2247 (2017).

%CPI

\bibitem{cpi_review}
F. Di Lena, F. V. Pepe, A. Garuccio, and M. D'Angelo, ``Correlation Plenoptic Imaging: An Overview,'' Appl. Sci. \textbf{8}, 1958 (2018).

\bibitem{cpi_prl}
M.~D'Angelo, F.~V. Pepe, A.~Garuccio, and G.~Scarcelli, ``Correlation Plenoptic Imaging,'' Phys. Rev. Lett. \textbf{116}, 223602 (2016).

\bibitem{cpi_qmqm}
F.~V. Pepe, G.~Scarcelli, A.~Garuccio, and M.~D'Angelo, ``Plenoptic imaging with second-order correlations of light,'' Quantum Meas. Quantum Metrol. \textbf{3}, 20 (2016).

\bibitem{cpi_jopt}
F. V. Pepe, O. Vaccarelli, A. Garuccio, G. Scarcelli, and M. D'Angelo, ``Exploring plenoptic properties of correlation imaging with chaotic light,'' J. Opt. \textbf{19}, 114001 (2017).

\bibitem{cpi_exp}
F. V. Pepe, F. Di Lena, A. Mazzilli, E. Edrei, A. Garuccio, G. Scarcelli, and M. D'Angelo, ``Diffraction-Limited Plenoptic Imaging with Correlated Light,'' Phys. Rev. Lett. \textbf{119}, 243602 (2017).

\bibitem{cpi_technologies}
F.~V. Pepe, F.~Di~Lena, A.~Garuccio, G.~Scarcelli, and M.~D'Angelo, ``Correlation plenoptic imaging with entangled photons,'' Technologies--Open Access Multidisciplinary Engineering Journal \textbf{4}, 17 (2016).

%GHOST IMAGING

\bibitem{gatti}
A.~Gatti, E.~Brambilla, M.~Bache, and L.~A. Lugiato, ``Ghost imaging with thermal light: comparing entanglement and classical correlation,'' Phys. Rev. Lett. \textbf{93}, 093602 (2004).

\bibitem{bennink}
R. S. Bennink, S. J. Bentley, R. W. Boyd, and J. C. Howell, ``Quantum and Classical Coincidence Imaging", Phys. Rev. Lett. \textbf{92}, 033601 (2004).

\bibitem{valencia}
A.~Valencia, G.~Scarcelli, M.~D'Angelo, and Y.~Shih, ``Two-photon imaging with thermal light,'' Phys. Rev. Lett. \textbf{94}, 063601 (2005).

\bibitem{scarcelliPRL}
G.~Scarcelli, V.~Berardi, and Y.~Shih, ``Can two-photon correlation of chaotic light be considered as correlation of intensity fluctuations?,'' Phys. Rev. Lett. \textbf{96}, 063602 (2006).

\bibitem{devaux}
F. Devaux, K. P. Huy, S. Denis, E. Lantz, and P.-A. Moreau, ``Temporal ghost imaging with pseudothermal speckle light,'' J. Opt. \textbf{19}, 024001 (2017).

\bibitem{laserphys}
M.~D'Angelo and Y.~Shih, ``Quantum imaging,'' Laser Phys. Lett. \textbf{2}, 567 (2005).

\bibitem{shapiro_review}
J. H. Shapiro and R. W. Boyd, ``The physics of ghost imaging'', Quantum Inf. Process. \textbf{11}, 949 (2012).

%SNR IN GHOST IMAGING

\bibitem{gatti_coh}
A. Gatti, M. Bache, D. Magatti, E. Brambilla, F. Ferri, and L. A. Lugiato, ``Coherent imaging with pseudo-thermal incoherent light,'' J. Mod. Opt. \textbf{53}, 739 (2006).

\bibitem{erkmen}
B. I. Erkmen and J. H. Shapiro, ``Signal-to-noise ratio of Gaussian-state ghost imaging,'' Phys. Rev. A \textbf{79}, 023833 (2009).

\bibitem{osullivan}
M. N. O'Sullivan, K. W. C. Chan, and R. W. Boyd, ``Comparison of the signal-to-noise characteristics of quantum versus thermal ghost imaging,'' Phys. Rev. A \textbf{82}, 053803 (2010).

\bibitem{brida_pra}
G. Brida, M. V. Chekhova, G. A. Fornaro, M. Genovese, E. D. Lopaeva, and I. Ruo Berchera, ``Systematic analysis of signal-to-noise ratio in bipartite ghost imaging with classical and quantum light,'' Phys. Rev. A \textbf{83}, 063807 (2011).


%COMPRESSIVE SENSING AND DIFFERENTIAL GI

\bibitem{katz}
O. Katz, Y. Bromberg, and Y. Silberberg, ``Compressive ghost imaging,'' Appl. Phys. Lett. \textbf{95}, 131110 (2009).

\bibitem{welsh}
S. S. Welsh, M. P. Edgar, P. Jonathan, B. Sun, and M. J. Padgett, ``Multi-wavelength compressive computational ghost imaging,'' Proc. SPIE \textbf{8618}, 86180I-1 (2013).

\bibitem{ferri_dgi}
F. Ferri, D. Magatti, L. A. Lugiato, and A. Gatti, ``Differential Ghost Imaging,'' Phys. Rev. Lett \textbf{104}, 253603 (2010).


\bibitem{mandel}
L. Mandel, E. Wolf, {\em Optical Coherence and Quantum Optics} (Cambridge University Press, Cambridge, 1995).

\bibitem{isserlis}
L. Isserlis, ``On a formula for the product-moment coefficient of any order of a normal frequency distribution in any number of variables,'' Biometrika \textbf{12}, 134 (1918).

\bibitem{goodman}
J. W. Goodman, \textit{Introduction to Fourier Optics} (McGraw-Hill, New York, 1996).

%X RAYS

\bibitem{pelliccia}
D. Pelliccia, A. Rack, M. Scheel, V. Cantelli, and D. M. Paganin, ``Experimental X-Ray Ghost Imaging,'' Phys. Rev. Lett. \textbf{117}, 113902 (2016).

\bibitem{schneider}
R. Schneider \textit{et al.}, ``Quantum imaging with incoherently scattered light from a free-electron laser,'' Nat. Phys. \textbf{14}, 126 (2018).

%SUB-SHOT-NOISE

\bibitem{brida_nat}
G. Brida, M. Genovese, and I. Ruo Berchera, ``Experimental realization of sub-shot-noise quantum imaging,'' Nat. Photonics \textbf{4}, 227 (2010).

\bibitem{meda}
A. Meda, E. Losero, N. Samantaray, F. Scafirimuto, S. Pradyumna, A. Avella, I. Ruo-Berchera, and M. Genovese, ``Photon-number correlation for quantum enhanced imaging and sensing,'' J. Opt. \textbf{19}, 094002 (2017).

\bibitem{samantaray}
N. Samantaray, I. Ruo-Berchera, A. Meda, and M. Genovese, ``Realization of the first sub-shot-noise wide field microscope,'' Light Sci. Appl. \textbf{6}, e17005 (2017).





\end{thebibliography}
\end{document}